\documentclass[preprint,showpacs,preprintnumbers,amsmath,amssymb]{revtex4}
\usepackage{booktabs}
\usepackage{mathrsfs}
\usepackage{epsfig}
\usepackage{graphicx}
\usepackage{dcolumn}
\usepackage{bm}
\usepackage{amsmath}
\usepackage{slashed}       
\usepackage{amssymb}
\usepackage{amsfonts}

\let\jnfont=\rm
\def\NPB#1,{{\jnfont Nucl.\ Phys.\ B }{\bf #1},}
\def\PLB#1,{{\jnfont Phys.\ Lett.\ B }{\bf #1},}
\def\EPJC#1,{{\jnfont Eur.\ Phys.\ Jour.\ C }{\bf #1},}
\def\PRD#1,{{\jnfont Phys.\ Rev.\ D }{\bf #1},}
\def\PRL#1,{{\jnfont Phys.\ Rev.\ Lett.\ }{\bf #1},}
\def\MPLA#1,{{\jnfont Mod.\ Phys.\ Lett.\ A }{\bf #1},}
\def\JPG#1,{{\jnfont J.\ Phys.\ G}{\bf #1},}
\def\CTP#1,{{\jnfont Commun.\ Theor.\ Phys.\ }{\bf #1},}
\def\ZPC#1,{{\jnfont Z.\ Phys.\ C }{\bf #1},}
\def\JHEP#1,{{\jnfont JHEP \ }{\bf #1},}
\def\lsim{\raise0.3ex\hbox{$<$\kern-0.75em\raise-1.1ex\hbox{$\sim$}}}
\def\gsim{\raise0.3ex\hbox{$>$\kern-0.75em\raise-1.1ex\hbox{$\sim$}}}

\begin{document}
\preprint{\parbox{1.2in}{\noindent arXiv:}}

\title{Higgs-strahlung production process $e^+ e^- \to Z h$ \\ at the future Higgs factory in the Minimal Dilaton Model}

\author{Junjie Cao$^{1,2}$, Zhaoxia Heng$^1$, Dongwei Li$^1$, Liangliang Shang$^1$, Peiwen Wu$^3$}

\affiliation{
  $^1$   Department of Physics,
        Henan Normal University, Xinxiang 453007, China \\
  $^2$  Center for High Energy Physics, Peking University,
       Beijing 100871, China \\
  $^3$   State Key Laboratory of Theoretical Physics,
      Institute of Theoretical Physics, Academia Sinica, Beijing 100190, China
      \vspace{1cm}}

\begin{abstract}
We investigate the Higgs-strahlung production process $e^+ e^- \to Z h$ at the future Higgs factory
such as TLEP by including radiative corrections in the Minimal Dilaton Model (MDM),
which extends the SM by one singlet scalar called dilaton.
We consider various theoretical and experimental constraints on the model, and perform fits to
the Higgs data taken from ATLAS, CMS and CDF+D0. Then for the 1$\sigma$ surviving samples,
we calculate the MDM predictions on the inclusive production rate $\sigma(e^+e^-\to Zh)$ at the 240-GeV
Higgs factory, and also the signal rates of $e^+e^-\to Zh$ with the Higgs boson decaying to $b\bar b$
and $\gamma\gamma$. We have following observations:
(1) In the heavy dilaton scenario, the deviation of $\sigma(e^+e^-\to Zh)$ from its
SM prediction can vary from $-15\%$ to $85\%$, which mainly arises from the modification of the tree-level $hZZ$ coupling
and also the radiative correction induced by possibly large Higgs self-couplings. (2) The processes $e^+e^-\to Zh$ at
the Higgs factory and $pp\to hh$ at 14-TeV LHC
are complementary in limiting the MDM parameter space. Requiring the deviation of $\sigma(e^+e^-\to Zh)$
from its SM prediction to be less than $1\%$ and that of $\sigma(p p \to h h)$ to be less than $50\%$,
$\tan \theta_S$  in the MDM will be limited to be $-0.1<\tan\theta_S<0.3$, and the deviations of the
signal rates are constrained to be $|R_{b\bar b}|<2\%$ and $|R_{\gamma\gamma}|<7\%$. Especially, the Higgs self-coupling
normalized to its SM prediction is now upper bounded by about 4.
(3) In the light dilaton scenario, the deviation of $\sigma(e^+e^-\to Zh)$ may reach $-7\%$,  and
requiring its size to be less than $1\%$ will result in $0<\tan\theta_S<0.1$, and $-10\% < R_{b\bar b}, R_{\gamma\gamma} < 1\%$.
\end{abstract}


\maketitle

\section{Introduction}
In July 2012, the discovery of a new boson with mass around 125 GeV at the CERN Large Hadron Collider (LHC) \cite{1207atlas,1207cms}
marked a great triumph in the history of particle physics. With the growingly accumulated data,
the properties of this newly discovered boson are in excellent agreement with those of the Higgs boson predicted
by the Standard Model (SM), including the further measurements of its spin and
parity quantum numbers \cite{spin-atlas,spin-cms}.
However, up to now, there is no evidence to establish whether the Higgs sector contains
only one Higgs doublet. Instead, the Higgs-like resonance with mass
about 125 GeV can also be well explained in many new physics models, such as
low energy supersymmetric models \cite{125-MSSM,125-NMSSM} and
the dilaton models \cite{dilaton-125}.

So far various Higgs couplings to SM fermions and vector bosons based on the current LHC data still have large uncertainties. Taking the $hZZ$ coupling as an example, the measured value normalized to its SM prediction is $1.43\pm 0.33({\rm stat})\pm 0.17 ({\rm syst})$ for ATLAS result and $0.92\pm 0.28$ for CMS result \cite{Higgs_working_group_report_2013}. Nevertheless, at the future High Luminosity LHC (HL-LHC) with $300\, {\rm fb}^{-1}$ ($3000\, {\rm fb}^{-1}$) integrated luminosity, the precision of the $C_{hZZ}$ measurement is expected to reach $4-6\%$ ($2-4\%$) \cite{Higgs_working_group_report_2013}. Compared with the hadron collider, the future $e^+e^-$ collider may have a stronger capability in the $C_{hZZ}$ measurement through the Higgs-strahlung production $e^+e^-\to Zh$. For example, at the  proposed
International Linear Collider (ILC) with collision energy up to $1 {\rm TeV}$ and luminosity up to $1000 fb^{-1}$, the precision may be improved to be near $0.5\%$\cite{Higgs_working_group_report_2013}. And an even more remarkable precision of $0.05\%$ may be achieved at the recently proposed Triple-Large Electron-Positron Collider (TLEP)\cite{Higgs_working_group_report_2013}, which is a new circular $e^+e^-$ collider operated at $\sqrt{s}=$ 240 GeV with $10^4 fb^{-1}$ integrated luminosity\cite{TLEP}.

The story of the Higgs self-coupling, however, is quite different. By now such a coupling has basically not been constrained by
the current Higgs data, while on the other hand, it can be quite large in some new physics models such as
the Minimal Dilaton Model (MDM) \cite{dilaton1,dilaton2,dilaton3}. Obviously,
the next important task of experimentalists is to determine the coupling size as precise as possible, which is essential
in reconstructing the Higgs potential and consequently determining the mechanism of the electro-weak symmetry breaking.
At both the LHC and the ILC, the Higgs self-coupling can be
measured directly through the Higgs pair production \cite{QCD-pphh,SUSY-Higgs pair,Other-Higgs pair}.
The recent studies suggest that a precision of $50\%$ for the coupling can be obtained through $pp\to hh\to bb\gamma\gamma$ at the HL-LHC with
an integrated luminosity of $3000\, {\rm fb}^{-1}$ \cite{Higgs_working_group_report_2013,Higgs-self-coupling}, and it may be further improved
to be around $13\%$ at the ILC with collision energy up to $1 {\rm TeV}$\cite{Higgs_working_group_report_2013}.

One interesting feature of the Higgs-strahlung production $e^+e^-\to Zh$ is that, while at tree level its rate is solely determined by the
$C_{hZZ}$ coupling, at loop level the rate also depends on the Higgs self-coupling and may be significantly altered by such a
coupling.  This brings us the possibility that apart from the direct Higgs pair production,
the Higgs self-coupling may also be measured indirectly from the process $e^+e^-\to Zh$
with the $e^+e^-$ collision energy below the di-Higgs threshold.
As shown in \cite{McCullough:2013rea},  given that  the inclusive cross section  $\sigma(e^+e^-\to Zh)$ is measured
with a precision of 0.4\% at the TLEP\cite{TLEP}, the Higgs self-coupling may be constrained to an accuracy of $28\%$.

In this work we take the MDM as an example to investigate the Higgs-strahlung production $e^+e^-\to Zh$ by including radiative corrections.
We  scan the MDM parameters by considering various experimental and theoretical constraints.  Then for the surviving samples
we calculate their predictions on $\sigma(e^+e^-\to Zh)$ at the 240-GeV TLEP, and investigate
to what extent the parameters will be constrained given the future precision of the cross section measurement. Noting that more
observables will be helpful to further limit the parameter space, we also perform a study on the signals of the Higgs-strahlung  production
with the Higgs boson decaying to $\gamma\gamma$ or $b\bar b$. We note that similar study has been done in supersymmetric
models \cite{Hu:2014eia}.

This work is organized as follows. In Sec. II, we briefly review the MDM and experimental and theoretical constraints on it. Then we calculate $\sigma(e^+e^-\to Zh)$ by including radiative corrections and discuss the capability of the
 Higgs factory to determine the model parameters in Sec. III. Finally, we summarize our conclusions in Sec. IV.

\section{The Minimal Dilaton Model}
The MDM is an extension of the SM by introducing a
linearized singlet dilaton field $S$ and a vector-like top partner $T$ with the same quantum
number as the right-handed top quark. The low energy effective Lagrangian of the MDM
is given by \cite{dilaton1,dilaton2,dilaton3}
\begin{eqnarray}
  \mathcal{L} &=& \mathcal{L}_{\rm SM}-\frac{1}{2}\partial_\mu S\partial^\mu S
  -{m_S^2\over2}S^2-{\lambda_S\over 4!}S^4-{\kappa\over2}S^2\left|H\right|^2
		-m_H^2\left|H\right|^2-{\lambda_H\over4}\left|H\right|^4  \nonumber \\
	&&   -\bar{T} \left(\slashed{D}+\frac{M}{f}S\right)T
		-\left[y'\overline T_R(q_{3L}\cdot H)+\text{h.c.}\right],    \label{Lagrangian}
\end{eqnarray}
where $\mathcal{L}_{\rm SM}$ is the part of the SM Lagrangian without the Higgs potential, $M$ represents the scale of a certain
strong dynamics in which the fields $T$ and $S$ are involved, $q_{3L}$ is the $SU(2)_L$ left-handed quark doublet of the third generation, and
$M_H$, $M_S$, $\lambda_S$, $\kappa$ and $\lambda_H$ are all free parameters describing the new Higgs potential.
The singlet dilaton field $S$ and the doublet Higgs field $H$ will mix with each other, which can be parameterized by the Higgs-dilaton mixing angle $\theta_S$ as
\begin{eqnarray}
  S &=& f+h\sin\theta_S+s\cos\theta_S,\nonumber\\
  H &=& \left(\begin{array}{c}
          \phi^+ \\
          \frac{1}{\sqrt{2}} (v+h\cos\theta_S-s\sin\theta_S+i\phi^0)
        \end{array} \right)
\end{eqnarray}
with $f$ and $v=246$ GeV being the vacuum expectation values (vev) of $S$ and $H$ respectively,
$h$ and $s$ denoting the mass eigenstates of the Higgs boson and the dialton, and $\phi^0$ and $\phi^+$ representing the Goldstone bosons.
Similarly, $q_{3L}^u$ and $T$ will mix to form mass eigenstates $t$ and $t^\prime$ so that
\begin{eqnarray}
q_{3L}^u &=& \cos \theta_L t_L + \sin \theta_L t^\prime_L, \nonumber \\
T_L &=& - \sin \theta_L t_L + \cos \theta_L t^\prime_L.
\end{eqnarray}

\vspace{-0.1cm}

If $\theta_S$, $f$ and physical masses $m_h$, $m_s$ are taken as the input of the theory,
one can re-express the dimensionless parameters $\lambda_S$, $\kappa$ and $\lambda_H$ as follows\cite{dilaton3}
\begin{eqnarray}
\lambda_S &=& \frac{3|m_h^2 - m_s^2|}{2f^2} ~\left[\frac{m_h^2 + m_s^2}{|m_h^2 - m_s^2|}
        -\rm Sign{(\sin2\theta_S)} \cos2\theta_S \right],\nonumber\\
\kappa &=& \frac{|m_h^2 - m_s^2|}{2 f v} |\sin2\theta_S|, \nonumber\\
\lambda_H &=& \frac{|m_h^2 - m_s^2|}{v^2} ~\left[\frac{m_h^2 + m_s^2}{|m_h^2 - m_s^2|}
        + \rm Sign{(\sin2\theta_S)} \cos2\theta_S \right].
\label{lambda}
\end{eqnarray}
In this case, the trilinear interactions among $h$, $s$, $\phi^0$ and $\phi^\pm$ are given by
\begin{eqnarray}
C_{hhh} = v
&\big [&~\frac{3}{2} \lambda_H\cos^3\theta_S +\lambda_S\eta^{-1}\sin^3\theta_S
+3\kappa(\cos\theta_S\sin^2\theta_S +\eta^{-1}\cos^2\theta_S\sin\theta_S) ~\big],    \label{hhh}\\
C_{hss} = v &\big[& \kappa (\cos^3\theta_S + \eta^{-1} \sin^3\theta_S)
 + (\frac{3}{2} \lambda_H-2\kappa) \cos\theta_S \sin^2\theta_S  \nonumber\\
 && + \, \eta^{-1} (\lambda_S-2\kappa) \cos^2\theta_S \sin\theta_S ~\big],      \label{hss}  \\
C_{hhs} = v &\big[& \kappa (-\sin^3\theta_S + \eta^{-1} \cos^3\theta_S)
 - (\frac{3}{2} \lambda_H-2\kappa) \sin\theta_S \cos^2\theta_S   \nonumber\\
 && + \, \eta^{-1} (\lambda_S-2\kappa) \sin^2\theta_S \cos\theta_S ~\big],      \label{hhs}  \\
C_{h\phi^0\phi^0}= v&(&\kappa\eta^{-1}\sin\theta_S+\frac{\lambda_H}{2}\cos\theta_S ~),   \\
C_{h\phi^+\phi^-}= v&(&\kappa\eta^{-1}\sin\theta_S+\frac{\lambda_H}{2}\cos\theta_S ~)
\end{eqnarray}
with $\eta \equiv \frac{v}{f}$, and the normalized couplings of $h$ and $s$ with
$Z$ or $\phi^0$ are given by
\begin{eqnarray}
C_{hZZ}/SM	&=& C_{hZ\phi^0}/SM=\cos\theta_S,~~
C_{sZZ}/SM	=C_{sZ\phi^0}/SM =-\sin\theta_S, \nonumber\\
C_{hhZZ}/SM &=& \cos^2\theta_S , ~~~
C_{hsZZ}/SM	=	-\cos\theta_S\sin\theta_S,~~~
C_{ssZZ}/SM = \sin^2\theta_S.
\label{hsz}
\end{eqnarray}

In the following we differentiate two scenarios according to the dilaton mass\cite{dilaton3}:
\begin{itemize}
  \item Heavy dilaton scenario:  $m_s>m_h$. An important feature of this scenario is that the trilinear Higgs self-coupling $C_{hhh}$ may be very large.
  \item Light dilaton scenario: $m_s < \frac{m_h}{2} \simeq$ 62 GeV. In this scenario,
  the Higgs exotic decay $h\to ss$ is open with a possible large branching ratio, while $C_{hhh}/SM$ is around at
  either $1$ or $0$.
\end{itemize}
For each parameter point of these scenarios, we impose the constraints similar to what we did in \cite{dilaton3}, which are given by
\begin{itemize}
\item[(1)] Vacuum stability of the scalar potential and absence of the Landau pole up to $1 {\rm TeV}$.
\item[(2)] Bounds from the search for Higgs-like scalar at LEP, Tevatron and LHC.
\item[(3)] $m_{t^\prime} \geq 1 {\rm TeV}$ as suggested by the LHC search for top quark partner\cite{ex-top-partner}
and constraints from the precision electroweak data\cite{dilaton1}. With this constraint, we have $\cos \theta_L > 0.97$ and
consequently $C_{htt}/SM \simeq \cos \theta_S$\cite{dilaton3}.
\item[(4)] Constraints from the measured Higgs properties. In implementing this constraint,
we use the combined data (22 sets) from ATLAS, CMS and CDF+D0 and perform a fit with the same method as that
in \cite{h-fit1,h-fit2,Cao:2013gba}. We obtained $\chi^2_{min}=$ 18.66 in the MDM, which is less
than $\chi^2$ for the SM ($\chi^2_{SM}=18.79$), and paid particular attention to $1 \sigma$ samples in the fit.
\end{itemize}

As shown in \cite{dilaton3}, parameter points satisfying the above constraints will predict
$\cos \theta_S > 0.92$, and $C_{h\bar{t^\prime}t^\prime}/C_{h\bar{t}t}^{SM} < 0.1$.
As will be seen below, this feature is beneficial for our analysis.

\section{Calculations and numerical results}

\begin{figure}[tp]
\includegraphics[width=11.5cm]{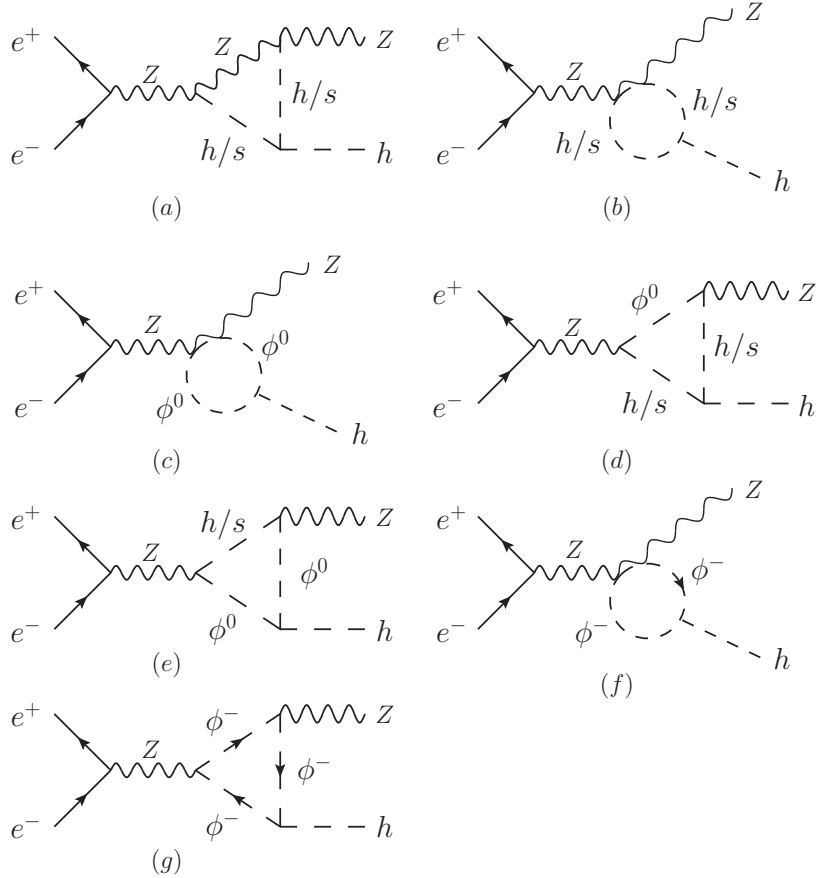}
\vspace{-0.5cm}
\caption{Feynman diagrams for the Higgs-strahlung production $e^+e^-\to Zh$
in the MDM with corrections
from the Higgs self-couplings at NLO in the Feynman-'t Hooft gauge.}
\label{fig1}
\end{figure}

In the SM, the radiative corrections to the Higgs-strahlung production process $e^+ e^- \to Z h$ come from the $Z$ boson self-energy, the vertex corrections
to $Ze^+e^-$, $h e^+ e^-$, $ZZh$ and $Z\gamma h$ interactions, and also box contributions\cite{Denner,Liu:2013cla}. 
The full calculation of these corrections is quite
complex (e.g. more than sixty diagrams need to be caculated) and it was shown recently that the total weak correction is
$5\%$ for $m_h=125 {\rm GeV}$ and $\sqrt{s}= 240 {\rm GeV}$\cite{Zh-radiative}.
About the corresponding corrections in the MDM, we have following observations
\begin{itemize}
\item Although the contribution induced by the Higgs self-coupling is only $2\%$ in the SM\cite{Zh-radiative}, it is potentially large in the MDM
 since the self-couplings among the scalars may be greatly enhanced\cite{dilaton3}. In this work, we will focus on such a contribution.
\item The correction mediated by $t^\prime$ quark can be  safely neglected since $t^\prime$ is heavy and meanwhile $C_{h\bar{t}^\prime t^\prime}$ is relatively small.
\item For loops that involves the $s Z Z$ interaction and meanwhile do not involve possible large self-couplings among the scalars,  their contributions are  negligible since the dilaton is highly singlet dominated.
\item For the rest contributions, they can be obtained from the corresponding SM results in \cite{Denner} by scaling with a factor of either $\cos^3 \theta_S$ or $\cos \theta_S$. We find by detailed calculation that the size of the former contribution, i.e. that obtained by the scaling factor of $\cos^3 \theta_S$,  is $-0.4\%$ in the SM, and the latter contribution is $3.4\%$.
\end{itemize}
Based on these observations, we conclude that the deviation of the inclusive production rate $\sigma(e^+ e^- \to Z h)$
from its SM prediction, which is generally called
genuine new physics contribution, is given by
\begin{eqnarray}
  R &\equiv& \frac{\sigma_{\rm MDM}^{\rm LOOP}(e^+e^-\to Zh)-\sigma_{\rm SM}^{\rm LOOP}(e^+e^-\to Zh)}{\sigma_{\rm SM}^{\rm 0}(e^+e^-\to Zh)}  \nonumber \\
    &\simeq& \cos^2 \theta_S + 0.034 \cos^2 \theta_S - 0.004 \cos^4 \theta_S  + \frac{\delta \sigma_{\rm MDM}^{\rm scalar}(e^+e^-\to Zh)}{\sigma_{\rm SM}^{\rm 0}(e^+e^-\to Zh)}
    - 1.05 \nonumber \\
    &\simeq & 1.03 \cos^2 \theta_S  + \frac{\delta \sigma_{\rm MDM}^{\rm scalar}(e^+e^-\to Zh)}{\sigma_{\rm SM}^{\rm 0}(e^+e^-\to Zh)}  + 0.001 \sin^2 2 \theta_S - 1.05
\end{eqnarray}
where $\sigma_{\rm MDM}^{\rm LOOP}$ and $\sigma_{\rm SM}^{\rm LOOP}$ are the cross sections at one loop level in the MDM and the SM respectively,
$\sigma_{\rm SM}^{\rm 0}$ is the SM prediction on the cross section at tree level, and $\delta \sigma_{\rm MDM}^{\rm scalar} $ denotes the correction induced by
the self-couplings among the scalars with the corresponding diagrams given in Fig.\ref{fig1}. Note that the first term on the right hand of the second equation
corresponds to the tree-level contribution, which differs from its SM prediction due to the modified $hZZ$ coupling by a factor $\cos \theta_S$. Also note that the constraints
 have required $\cos \theta_S > 0.92$, so the deviation
$R$ mainly arises from the modification of the tree-level $h ZZ$ coupling and $\delta \sigma_{\rm MDM}^{\rm scalar} $.

In this work, we take $m_Z=91.19 {\rm GeV},\, \alpha=1/128$ \cite{PDG} and $m_h = 125 {\rm GeV}$, and fix the $e^+e^-$ collision energy at 240 GeV.
We obtained $\sigma^0_{SM} (e^+e^-\to Zh) = 236 fb$, which is in accordance with the result in \cite{TLEP}.
In our calculations of $\delta \sigma_{\rm MDM}^{\rm scalar} $,  we adopt the Feynman-'t Hooft gauge, and therefore the diagram involving the Goldstone
fields must be considered. Moreover, we note from Fig.\ref{fig1} that, except for the dilaton mass, the masses of the particles in the loops are fixed,
and meanwhile, since the dilaton coupling with $Z$ boson is very small due to its singlet-dominated nature, its induced contribution should
be relatively small if $C_{hss}$ or $C_{hhs}$ is not much larger than $C_{hhh}$.  These features imply that $\delta \sigma_{\rm MDM}^{\rm scalar}$ or $R$ can be expressed
in a semi-analytic way, which is given by
\begin{eqnarray}
  R &\simeq & 1.03 \cos^2 \theta_S  + 0.02\times\cos\theta_S\times\frac{C_{hhh}}{\rm SM}
            + 0.000146 \times (\frac{C_{hhh}}{\rm SM})^2 \nonumber \\ && + 0.001 \sin^2 2 \theta_S -1.05.  \label{deviation}
\end{eqnarray}
In above equation, the second term on the right side reflects the interference between the tree-level contribution and the correction from the self-couplings,
the third represents the pure self-coupling contribution which can not be neglected if $C_{hhh}/SM \gg 1$, and the fourth term can be safely neglected given $\cos \theta_S > 0.92$. For the results presented below, we obtain the value of $\delta \sigma_{\rm MDM}^{\rm scalar} $ by exact calculation, and we
checked that for nearly all
the surviving samples, Eq.(\ref{deviation}) is a good approximation.

\begin{figure}[tp]
\includegraphics[width=11.5cm]{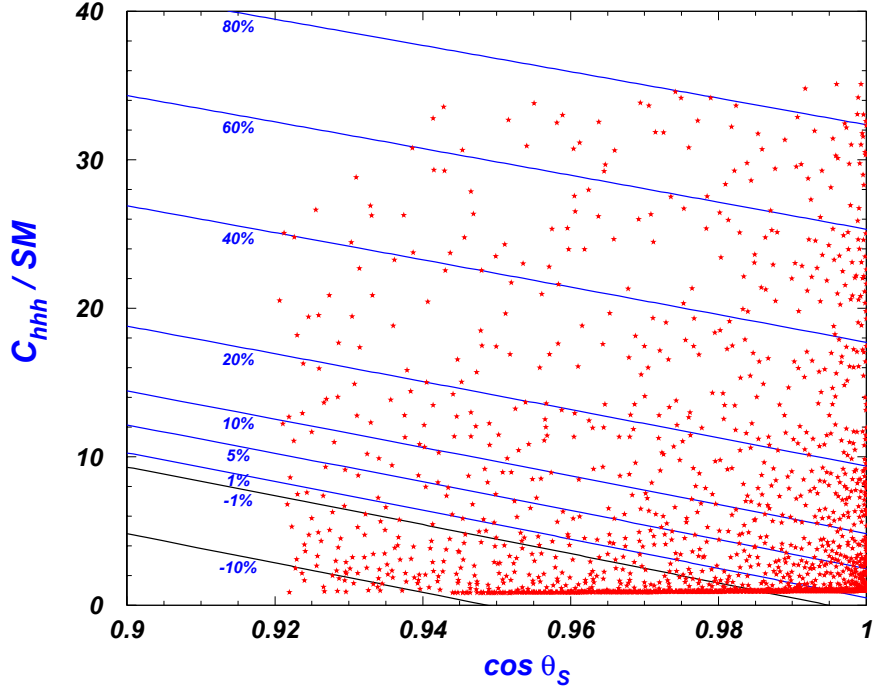}
\vspace{-0.5cm}
\caption{The scatter plot of the $1\sigma$ samples in the heavy dilaton scenario, projected on the
plane of $C_{hhh}/SM$ versus $\cos\theta_S$. The lines denote various specific values of the deviation $R$ calculated from Eq.(\ref{deviation}).}
\label{fig2}
\end{figure}

\subsection{Numerical results in the heavy dilaton scenario}
For the heavy dilaton scenario, we consider the constraints listed in Sect. II  and scan the relevant parameters
in the following ranges like what  we did in \cite{dilaton3}
\begin{eqnarray}
 1\leq \eta^{-1} <10, ~~130~{\rm GeV} <m_s< 1~{\rm TeV},
  ~~ |\tan\theta_S|<2,~~ 1{\rm TeV} < m_{t^\prime} < 3 {\rm TeV}.   \label{scan-ranges}
\end{eqnarray}
Then we investigate the properties of the $1 \sigma$ samples, which satisfy $\chi^2 -\chi^2_{min} \leq 2.3$.

\begin{figure}[tp]
\includegraphics[width=11.5cm]{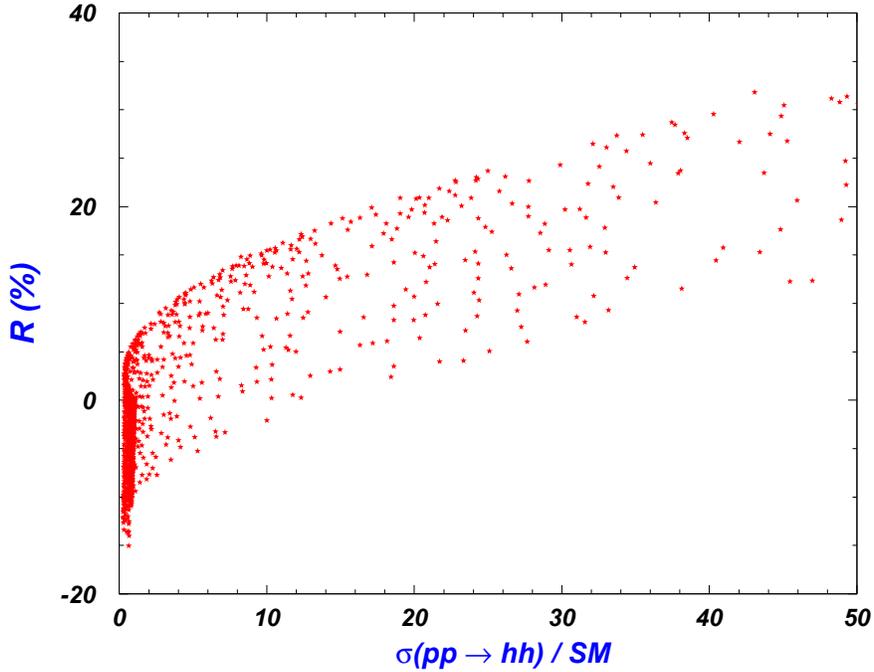}
\vspace{-0.5cm}
\caption{Same as Fig.\ref{fig2}, but projected on the
plane of $R$ versus the normalized cross section rate $\sigma(pp\to hh)/SM$ at the 14-TeV LHC.}
\label{fig3}
\end{figure}

In Fig.\ref{fig2} we project the $1\sigma$ samples on the plane of
$C_{hhh}/SM$ versus $\cos\theta_S$ and also show some lines corresponding to
specific values of $R$ calculated from Eq.(\ref{deviation}).
One can learn the following features:
\begin{itemize}
    \item Due to the small coefficients of the second and third terms in Eq.(\ref{deviation}), a given value of $R$ in Eq.(\ref{deviation}) corresponds to
    a very prolate ellipse on the whole plane of $C_{hhh}/SM$ versus $\cos\theta_S$ after neglecting the term proportional to $\sin^2 2 \theta_S$. For $\cos \theta_S > 0.92$,
        the  ellipse curves turn out to be nearly straight lines in Fig.\ref{fig2}.
    \item As indicated by Eq.(\ref{deviation}), the tree-level modification of the $hZZ$ coupling is to decrease the inclusive rate, while the effect of the correction induced
    by the self-couplings is to increase the rate. For the $1\sigma$ samples considered, the deviation $R$ varies from $-15\%$ to $85\%$. Such possible large deviation
    is due to a large uncertainty in determining the $hZZ$ coupling from current Higgs data as well as currently a very weak constraint on the self-couplings.

    Obviously, if $R$ is moderately large, two loop or higher order corrections should also be taken into account.
\end{itemize}

\begin{figure}[tp]
\includegraphics[width=11.5cm]{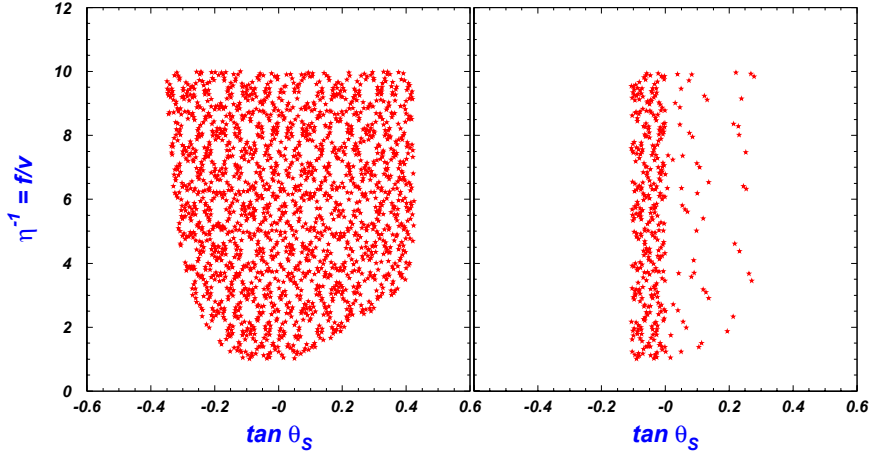}
\vspace{-0.5cm}
\caption{Same as Fig.\ref{fig2}, but projected on the plane of
$\eta^{-1}=f/v$ versus $\tan\theta_S$. The left panel shows all the $1\sigma$ samples, while samples in the right panel are further required to satisfy $|R|<1.0\%$ and $|\sigma(pp\to hh)/SM-1| <50\%$.}
\label{fig4}
\end{figure}

With the upgraded energy and luminosity of the LHC, $C_{hhh}$ may be
measured directly through the Higgs pair production since it affects the production rate through the parton process $gg \to h^{\ast} \to h h$.
As shown in Fig.6 of \cite{dilaton3}, for $C_{hhh}/SM \gtrsim 2.5 $ in the heavy dilaton scenario of the MDM,
the normalized cross section  $\sigma(pp\to hh)/SM$ at the 14-TeV LHC increases monotonically as $C_{hhh}/SM$ becomes larger.
In order to compare the effect of the Higgs self-coupling at the LHC with that at the future Higgs factory,
we show the correlation of $\sigma(pp\to hh)$ at the 14-TeV LHC with $\sigma(e^+e^-\to Zh)$ at 240-GeV TLEP in Fig.\ref{fig3}.
This figure manifests that a $\sigma(pp\to hh)/SM$ of several tens usually corresponds to a $R$ larger than $5\%$. For example, in the case of
$\sigma(pp\to h h)/SM =40$, $R$ varies from $10\%$ to $30\%$. While on the other hand, even for $\sigma(pp\to h h)/SM \sim 1$, the size of $R$
may still be moderately large, changing from $-15\%$ to $5\%$. These features tell us that the processes $pp\to hh$ and $e^+e^-\to Zh$
are complementary in limiting the parameters of the MDM.

In order to investigate the capability of the future experiments to detect the parameter space of the MDM, we assume
a measurement precision of 1.0\% for $\sigma(e^+e^-\to Zh)$ at 240 GeV \cite{TLEP} and 50\% for $\sigma(pp\to hh)$ at 14 TeV\cite{Higgs_working_group_report_2013,Higgs-self-coupling}.
Then we show the allowed parameter region on $\eta^{-1}-\tan \theta_S$ plane in the right panel of Fig.\ref{fig4} by requiring
the $1 \sigma$ samples to further satisfy $|R|<1.0\%$ and $|\sigma(pp\to hh)/SM-1| <50\%$.  For comparison, we also show the $1\sigma$ samples
 in the left panel of Fig.\ref{fig4} without the requirement. Fig.\ref{fig4} indicates that $\tan\theta_S$ is allowed
to be within $-0.4<\tan\theta_S<0.4$ and $-0.1<\tan\theta_S<0.3$ before and after the requirement respectively. Furthermore,
we checked that, after imposing the requirement, the number of the $1\sigma$ samples in our random scan is reduced by more than $80\%$, and now
$C_{hhh}$ satisfies $0.98 \leq C_{hhh}/SM \leq 4.4$. This reflects the great power of the future experiments in limiting the MDM.

\begin{figure}[tp]
\includegraphics[width=12cm]{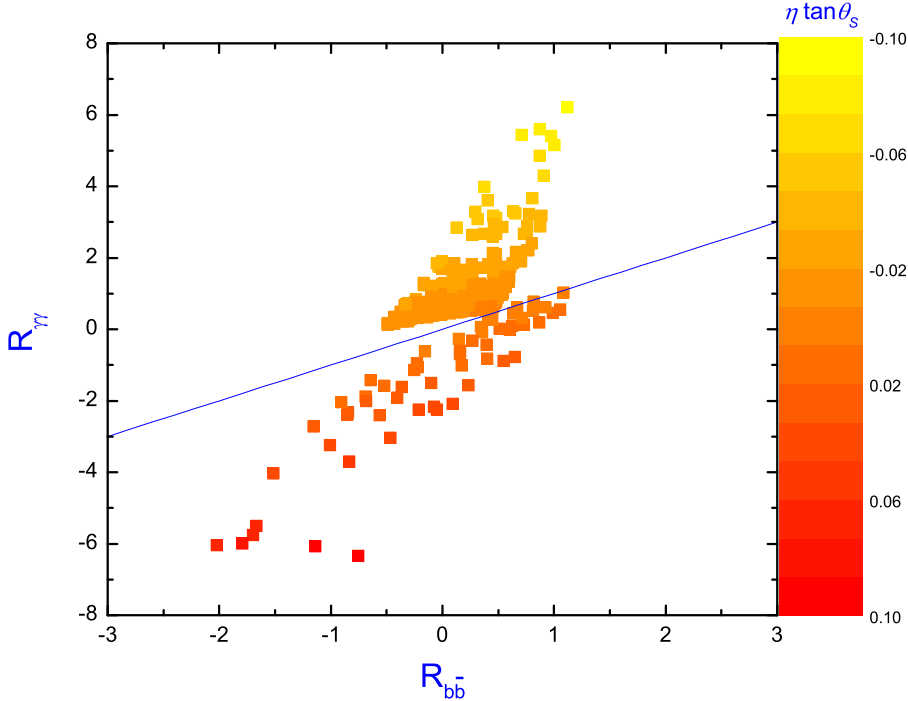}
\vspace{-0.5cm}
\caption{Samples in the right panel of Fig.\ref{fig4}, but projected on the plane of
$R_{\gamma\gamma}$ versus $R_{b\bar b}$, where dependence on $\eta\tan\theta_S$ is also shown. For clarity, we draw a blue line corresponding to $R_{b\bar{b}} = R_{rr}$.}
\label{fig5}
\end{figure}

Since the MDM parameters can still survive in a fairly wide region after considering the measurement of the
inclusive production rate at future Higgs factory, we need to consider more observables to limit the model. So we also investigate the signal rates
of  $e^+ e^- \to Z h \to Z b\bar b, Z \gamma\gamma$.  Similar to $R$, we define the deviations of the signal rates
from their SM predictions by
\begin{eqnarray}
  R_{b\bar b} &\equiv&  \frac{\sigma_{\rm MDM}^{\rm LOOP}(e^+e^-\to Zh) Br_{\rm MDM} (h \to b\bar b) - \sigma_{\rm SM}^{\rm LOOP}(e^+e^-\to Zh ) Br_{\rm SM}(h \to b\bar b) }{\sigma_{\rm SM}^{0}(e^+e^-\to Zh ) Br_{\rm SM} (h \to b\bar b)},  \nonumber\\
  R_{\gamma\gamma} &\equiv& \frac{\sigma_{\rm MDM}^{\rm LOOP}(e^+e^-\to Zh) Br_{\rm MDM} (h \to \gamma \gamma) - \sigma_{\rm SM}^{\rm LOOP}(e^+e^-\to Zh)  Br_{\rm SM} (h \to \gamma \gamma) }{\sigma_{\rm SM}^{\rm 0}(e^+e^-\to Zh )  Br_{\rm SM} ( h \to \gamma\gamma)}
       \label{R-signal}
\end{eqnarray}
where $Br_{MDM} (h \to b \bar{b})$ and $Br_{SM} ( h \to b \bar{b} )$ denote the branching ratio of $ h \to b \bar{b}$ in the MDM and the SM respectively, and similar notation is applied for $ h \to \gamma \gamma$.
In the heavy dilaton scenario, $R_{b\bar{b}}$ and $R_{\gamma\gamma}$ can be approximated by\cite{dilaton3}
\begin{eqnarray}
R_{b\bar b} &\simeq & (R+1.05) \cdot \frac{\cos^2\theta_S \Gamma_{SM}^{b\bar{b}}}{\cos^2\theta_S \Gamma_{SM}} \frac{\Gamma_{SM}}{\Gamma^{b\bar{b}}_{SM}}
 - 1.05 \simeq R,     \label{R-bb-heavy} \\
R_{\gamma\gamma} &\simeq &  (R + 1.05) \cdot (1-0.27\eta\tan\theta_S)^2 -1.05  \label{R-rr-heavy}
\end{eqnarray}
where $\Gamma_{SM}$ and $\Gamma_{SM}^{b\bar{b}}$  denote respectively the total width of the Higgs boson and the partial width of $h \to b \bar{b}$ in the SM.
Note that the above approximations are good only for a sufficiently large $R$, but anyhow, they are helpful to understand our results.
In Fig.\ref{fig5}, we project the samples in the right panel of Fig.\ref{fig4} on the plane of $R_{\gamma\gamma}$ versus $R_{b\bar b}$ for different
values of $\eta \tan \theta_S$. This figure indicates that $R_{b\bar b}$ is basically constrained in the range of $|R_{b\bar b}|<2\%$, while
$|R_{\gamma\gamma}|$ can maximally reach $7\%$. Considering that the expected precisions of measured $\sigma\cdot BR(h\to b\bar{b})$ and $\sigma\cdot BR(h\to \gamma\gamma)$ at 240-GeV TLEP can reach the level of $0.2\%$ and $3.0\%$ respectively \cite{Higgs_working_group_report_2013}, one can expect that by the measurement of the $b\bar{b}$ and $\gamma \gamma$ signal rates, one can get additional information about $\eta \tan \theta_S$ if the MDM is a correct theory.

\begin{figure}[tp]
\includegraphics[width=12cm]{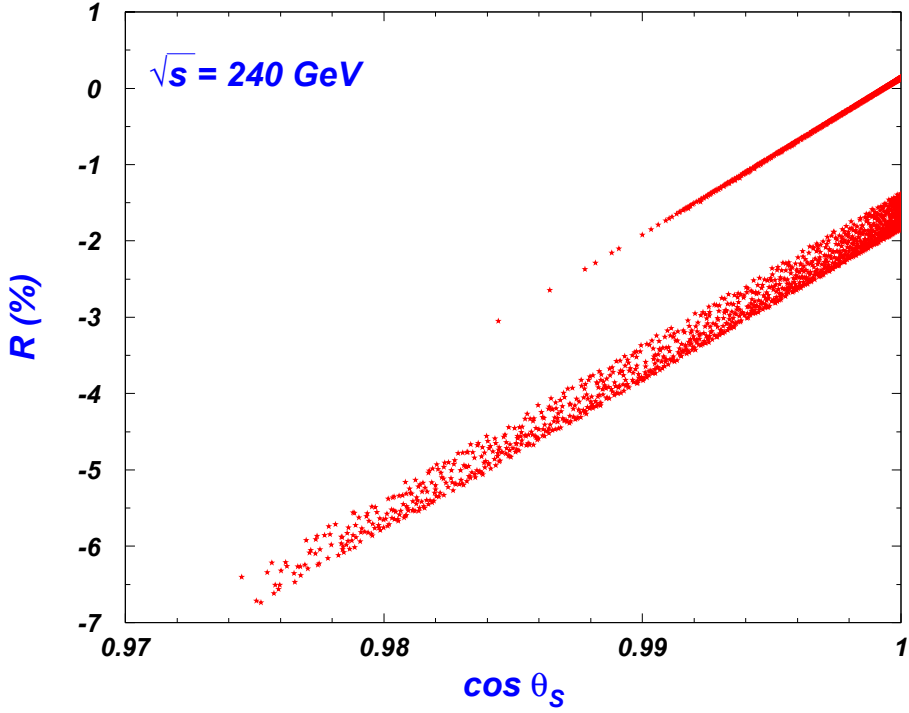}
\vspace{-0.5cm}
\caption{The scatter plot of the $1\sigma$ samples in the light dilaton scenario,
 projected on the plane of the deviation $R$ versus $\cos\theta_S$.}
\label{fig6}
\end{figure}

\subsection{Numerical results in the light dilaton scenario}

In the light dilaton scenario we scan following parameter ranges by considering the constraints listed in Sec. II
\begin{eqnarray}
 1\leq \eta^{-1} <10, ~~0~{\rm GeV} <m_s< 62~{\rm GeV},
  ~~ |\tan\theta_S|<2,~~ 1{\rm TeV} < m_{t^\prime} < 3 {\rm TeV},   \label{scan-range2}
\end{eqnarray}
and investigate the properties of the $1\sigma$ samples, which are now defined by $\chi^2-\chi^2_{min}\leq 1.0$\cite{dilaton3}.
Compared with the heavy dilaton scenario, the light dilaton scenario has two distinct features.
One is the Higgs exotic decay $h\to ss$ is open with a possible large branching ratio.
So this scenario is more tightly constrained by current Higgs data. And the other is the Higgs self-coupling
strength $C_{hhh}/SM$ is relatively small, around at either 1 or 0. As a result, the deviation
$R$ mainly comes from the modified $hZZ$ coupling, so $R \simeq \cos^2 \theta_S -1$.
In Fig.\ref{fig6} we project the $1\sigma$ samples on the plane of deviation $R$
versus $\cos\theta_S$. As expected, the size of the deviation $R$ monotonically decreases as $\cos\theta_S$ approach 1, and
it can maximally reach $7\%$.  This figure also shows that there are two separated regions of $R$. We checked that
it is due to the discontinuousness of $C_{hhh}/SM$, that is, the upper region corresponds to $C_{hhh}/SM \simeq 1$, while the lower region
corresponds to $C_{hhh}/SM \simeq 0$.

\begin{figure}[tp]
\includegraphics[width=12cm]{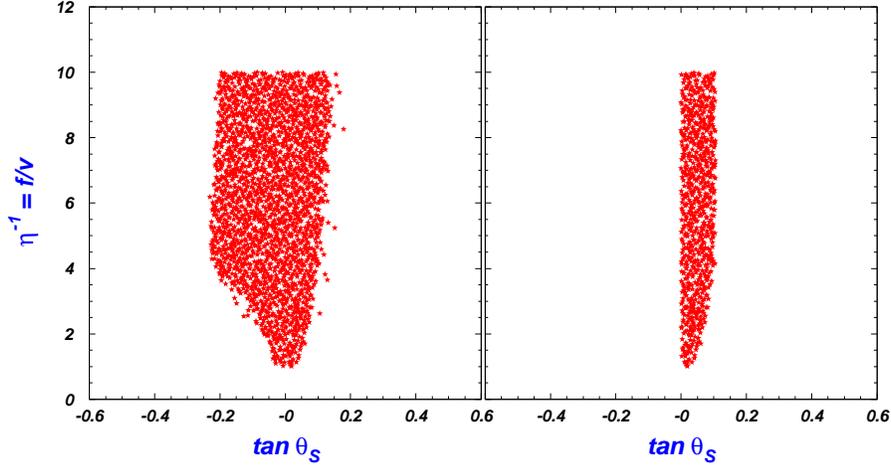}
\vspace{-0.5cm}
\caption{Scatter plot of the $1\sigma$ samples in the light dilaton scenario, projected on the plane of
$\eta^{-1}=f/v$ versus $\tan\theta_S$. The left panel shows all $1\sigma$ samples, while the right panel
shows samples further satisfying $|R|<1.0\%$.}
\label{fig7}
\end{figure}

Adopting the same analysis as Fig.\ref{fig4}, we show the $1\sigma$ samples projected on the plane of $\eta^{-1}=f/v$
versus $\tan\theta_S$ in Fig.\ref{fig7}, where the left panel shows all $1\sigma$ samples, while for comparison the right panel shows samples
that further satisfy the requirement $|R|<1.0\%$. Here we do not consider the deviation of $\sigma(pp\to hh)$ because it is very small in
the light dilation scenario \cite{dilaton3}. Fig.\ref{fig7} clearly shows that
the MDM parameter space in the light dilaton scenario is also strongly constrained resulting in $0<\tan\theta_S<0.1$,
in contrast with $-0.24 < \tan \theta_S < 0.2$  without the requirement of $|R| < 0.1$. Moreover, we checked that after the requirement, the number of the
$1\sigma$ samples in the left panel of Fig.\ref{fig7} is reduced by more than $70\%$.

\begin{figure}[tp]
\includegraphics[width=12cm]{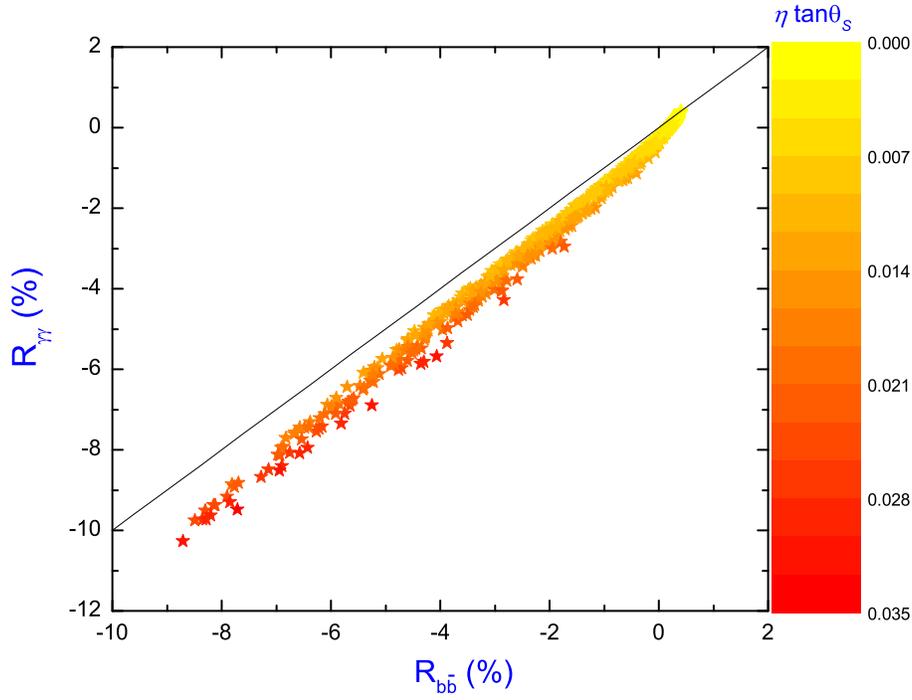}
\vspace{-0.5cm}
\caption{Same as Fig.\ref{fig6}, but projected on the plane of
$R_{\gamma\gamma}$ versus $R_{b\bar b}$, and also shows the dependence on $\eta\tan\theta_S$.}
\label{fig8}
\end{figure}

Similar to what we did in the heavy dilaton scenario, we also investigate the
signal deviations  $R_{b \bar b}$ and $R_{\gamma\gamma}$, which can now be expressed as
\begin{eqnarray}
R_{b\bar b} &\simeq & (R + 1.05) \frac{\cos^2\theta_S \Gamma^{b\bar{b}}_{SM}}{\cos^2 \theta_S \Gamma_{SM} +
\Gamma_{ss}} \frac{\Gamma_{SM}}{\Gamma^{b\bar{b}}_{SM}}  - 1.05 \nonumber \\
&\simeq & (R + 1.05) ( 1 - Br(h\to ss))  - 1.05  \label{R-bb-light} \\
R_{\gamma\gamma} &\simeq&  (R + 1.05) (1-0.27\eta\tan\theta_S)^2 (1 - Br(h\to ss))  -1.05,  \label{R-rr-light}
\end{eqnarray}
where $\Gamma_{ss}$ is the width of $h \to s s$ in the MDM.
In Fig.\ref{fig8} we show the relationship between $R_{\gamma\gamma}$ and $R_{b \bar b}$,
and their dependence on $\eta\tan\theta_S$.
From this figure we can see that $R_{\gamma\gamma}$ and $R_{b \bar b}$
follow a nearly linear correlation since now $\eta \tan \theta_S$ is very small, i.e. $|\eta \tan \theta_S| < 0.035$.
One can also see that even with the requirement $|R| < 1\%$,  $R_{b\bar b}$ and $R_{\gamma\gamma}$ may reach $-10\%$.
This is because the branching ratio of $h\to ss$ may still be moderate large under the constraint of current Higgs data.
Note that generally $|R_{\gamma\gamma}|$ is slightly larger than $|R_{b\bar b}|$, which can be understood by the positiveness of $\tan \theta_S$ in  Eq.(\ref{R-rr-light}).

\section{Summary and Conclusion}
In this work, we intend to investigate the capability of the future Higgs factory such as TLEP in detecting the parameter space of
the MDM, which  extends the SM by one singlet scalar called dilaton.
For this end, we calculate the Higgs-strahlung production process $e^+ e^- \to Z h$ at the future Higgs factory
by including radiative corrections in the model. We consider various theoretical and experimental constraints on the model, such as the vacuum stability, absence of Landau pole,
the electro-weak precision data and the LHC search for Higgs boson,
and perform fits to the Higgs data taken from ATLAS, CMS and CDF+D0. Then for the 1$\sigma$ surviving samples,
we investigate the MDM predictions on the inclusive production rate $\sigma(e^+e^-\to Zh)$ at the 240-GeV
Higgs factory, and also the signal rates of $e^+e^-\to Zh$ with the Higgs boson decaying to $b\bar b$
and $\gamma\gamma$. We have following observations:
(1) In the heavy dilaton scenario, the deviation of $\sigma(e^+e^-\to Zh)$ from its
SM prediction can vary from $-15\%$ to $85\%$, which mainly arises from the modification of the tree-level $hZZ$ coupling
and also the radiative correction induced by possibly large Higgs self-couplings. (2) The processes $e^+e^-\to Zh$ at
the Higgs factory and $pp\to hh$ at 14-TeV LHC
are complementary in limiting the MDM parameter space. Requiring the deviation of $\sigma(e^+e^-\to Zh)$
from its SM prediction to be less than $1\%$ and that of $\sigma(p p \to h h)$ to be less than $50\%$,
$\tan \theta_S$  in the MDM will be limited to be $-0.1<\tan\theta_S<0.3$, the deviations of the
signal rates are constrained to be $|R_{b\bar b}|<2\%$ and $|R_{\gamma\gamma}|<7\%$, and the Higgs self-coupling
normalized to its SM prediction is upper bounded by about 4.
(3) In the light dilaton scenario, the deviation of $\sigma(e^+e^-\to Zh)$ may reach $-7\%$,  and
requiring its size to be less than $1\%$ will result in $0<\tan\theta_S<0.1$, and $-10\% < R_{b\bar b}, R_{\gamma\gamma} < 1\%$.

\section*{Acknowledgement}
We thank Cheng Li, Jing-Ya Zhu and Li Lin Yang for helpful discussion at the early stage of this work.
This work is supported by the National Natural
Science Foundation of China (NNSFC) under grant No. 10775039, 11075045,
11275245, 10821504 and 11305050, by Program for New Century Excellent Talents in University,
by the Project of Knowledge Innovation Program (PKIP) of Chinese Academy of Sciences under grant No. KJCX2.YW.W10,
and by Specialized Research Fund for the Doctoral Program of Higher Education
with grant No. 20124104120001.

\end{document}